\documentclass[journal]{IEEEtran}
\ifCLASSINFOpdf
  \usepackage[pdftex]{graphicx}
  \graphicspath{{../} {../pdf/} {../jpeg/}}
  \DeclareGraphicsExtensions{.pdf,.jpeg,.png}
\else
  \usepackage[dvips]{graphicx}
  \graphicspath{{../} {../eps/}}
  \DeclareGraphicsExtensions{.eps}
\fi
\usepackage{amsmath,amssymb}
\usepackage{array}

\usepackage{multirow}
\usepackage{booktabs}
\usepackage{lipsum}
\usepackage{bm}
\usepackage{color}
\usepackage{url}

\usepackage{footnote}
\usepackage{tabularx}
\usepackage[flushleft]{threeparttable}

\begin{document}



\title{Applying Polynomial Chaos Expansion to Assess Probabilistic Available Delivery Capability for Distribution Networks with Renewables}


%

\author{Hao~Sheng,~\IEEEmembership{Member,~IEEE,} and~Xiaozhe~Wang,~\IEEEmembership{Member,~IEEE,}
\thanks{This work is partially supported by Natural Sciences and Engineering Research Council (NSERC) Discovery Grant, NSERC RGPIN-2016-04570.}
\thanks{H. Sheng and X. Wang are with the Department of Electrical and Computer Engineering, McGill University, Montr\'{e}al, QC, Canada. e-mail: shenghao@tju.edu.cn, xiaozhe.wang2@mcgill.ca}
}

%
%


\markboth{Accepted by IEEE Transactions on Power Systems for Future Publication}%
{Sheng \MakeLowercase{\textit{et al.}}: 
Applying Polynomial Chaos Expansion to Assess Probabilistic Available Delivery Capability for Distribution Networks with Renewables
}

%



\maketitle


\begin{abstract}
Considering the increasing penetration of renewable energy sources and electrical vehicles in utility distribution feeders, it is imperative to study the impacts of the resulting increasing uncertainty on the delivery capability of a distribution network. 
In this paper, probabilistic available delivery capability (ADC) is formulated for a general distribution network integrating various RES and load variations. To reduce the computational efforts by using conventional Monte Carlo simulations, we develop and employ a computationally efficient method to assess the probabilistic ADC, which combines the up-to-date sparse polynomial chaos expansion (PCE) and the continuation method. Particularly, the proposed method is able to handle a large number of correlated random inputs with different marginal distributions. 
Numerical examples in the IEEE 13 and IEEE 123 node test feeders are presented, showing that the proposed method can achieve accuracy and efficiency simultaneously. Numerical results also demonstrate that the randomness brought about by the RES and loads indeed leads to a reduction in the delivery capability of a distribution network. 
\end{abstract}

\begin{IEEEkeywords}
Available delivery capability (ADC), continuation method, Copula, distribution systems, Nataf transformation, polynomial chaos expansion, probabilistic continuation power flow (PCPF).
\end{IEEEkeywords}

%
\IEEEpeerreviewmaketitle


\section{Introduction}
%
%
%
%

\IEEEPARstart{T}{he} ever increasing integration of renewable energy sources (RES) (e.g., wind power and solar photovoltaic (PV)) and new forms of load demand (e.g., electric vehicles (EVs)) introduces more and more uncertainties to the utility distribution feeders. The resulting randomness in distribution system may affect the delivering capability of a distribution system. The concept of available delivery capability (ADC)  proposed in \cite{Sheng14a}, \cite{Chiang15} is used to describe the maximum power that can be delivered over the existing amount for which no thermal overloads, voltage violations, and voltage collapse occur. 
The minimum among the ADC subject to the thermal limits, the ADC subject to the voltage violations and the ADC subject to the voltage collapse is termed as the system overall ADC.
Nevertheless, the proposed ADC and the developed numerical method to solve it \cite{Sheng14a} did not account for the uncertainties coming from RES or EVs, which in turn are essential and stringent considering the constantly increasing integration of these components. In this paper, we propose a probabilistic framework to investigate the impact of uncertainties on the ADC of distribution networks.

The probabilistic analysis methods can be classified into two major categories: simulation methods and analytical methods. Monte Carlo simulation (MCS) \cite{Yu09} belonging to the first category is the most widely applied method  due to its simplicity, yet its high computational effort may prohibit its applications in real world, even with more efficient sampling methods such as Latin hypercube sampling \cite{Yu09}, importance sampling \cite{Huang11}, and Latin supercube sampling \cite{Hajian13}. 
Attempting to release the computational burden, analytical methods are developed utilizing mathematical approximations, among which the cumulant method \cite{Zhang04} and its extended versions \cite{Fan12} as well as point estimation method \cite{Su05}, \cite{Morales07} are the two representative ones. Nevertheless, the probability distributions of the power flow responses cannot be directly acquired from either cumulant methods or point estimation methods \cite{Ren16}. Another emerging method is the polynomial chaos expansion (PCE). 
The key idea of PCE is to represent the probabilistic response as a sum of orthogonal polynomial basis functions. It was firstly proposed by Wiener \cite{Wiener38} exclusively for Gaussian random variables using Hermite polynomial. Xiu \cite{Xiu02} further generalized the method to handle non-Gaussian random variables using a specific family of orthogonal polynomials. 
The popularity of the PCE method is mainly attributed to the following features: (i) It has strong mathematical basis; (ii) It can be readily integrated into existing deterministic analysis tools in a non-intrusive fashion. (ii) Accurate estimations for the probability distribution and the associated statistics can be obtained with low computational cost. 

PCE or generalized PCE (gPCE) has been applied in the context of power systems to study the probabilistic power flow \cite{Ren16} and the load margin problem \cite{Haesen09}, yet they suffer from `the curse of dimensionality' as the number of random inputs or the degree of polynomials increases. This issue has been addressed in \cite{Ni17} to solve the probabilistic power flow problem using a basis-adaptive sparse scheme that was originally developed in \cite{Blatman09}, \cite{UQLab14}. Our previous work \cite{Sheng17} combined the basis-adaptive sparse PCE (SPCE) and the continuation method to solve the ADC problem for general distribution networks with RES. 

However, little work to date has considered the cases in which a great number of random inputs with different marginal distributions co-exist and are highly correlated with each other. This situation, however, may be very common in power systems. Indeed, the stochastic characterizations for various RES are rather diverse. The wind speed may be described by Weibull distribution, Rayleigh distribution, and Gamma distribution \cite{Tuller85},\cite{Carta09}, whereas the solar radiation may follow Beta distribution or Weibull distribution \cite{Atwa10},\cite{Ettoumi02} depending on different locations and the time scales of interest. Furthermore, wind speeds,  solar radiations and load power at different sites within the same geographical area are typically dependent due to similar weather conditions and social customs. The challenges of handling such situation using the PCE method lie in the facts: 1) Finding the orthogonal polynomial basis for the distribution that a certain random input follows may be hard if the distribution is out of the regular list; 2) Obtaining the joint probability distribution of correlated random variables is difficult in practice. 

In this paper, we attempt to address these challenges of the conventional PCE and employ the modified PCE to assess probabilistic ADC for distribution networks with RES. Particularly, the first challenge aforementioned is resolved by employing the Stieltjes procedure which is able to construct a set of univariate polynomials orthogonal to any arbitrary probability distribution; the second challenge is overcome by leveraging on Copula and Nataf transformation to construct the joint distribution from available marginal distributions and correlation coefficient matrix. The contributions of the paper are as below:

\begin{itemize}
\item A mathematical formulation of probabilistic ADC problem based on continuation power flow is developed for distribution networks integrating RES generators and loads that may follow various marginal distributions. 
\item A computationally efficient yet accurate algorithm is proposed to evaluate the probabilistic ADC, which combines the sparse PCE method and continuation method. Particularly, for the PCE method: 
\begin{itemize}
\item Random variables with diverse marginal distributions can be accommodated using a set of mixed multivariate polynomial basis. 
\item Computational efficiency is improved by applying a flexible least angle regression (LAR)-based  adaptive sparse scheme. 
\item Correlation between random variables can be incorporated using the Nataf transformation. 
\end{itemize}
\item Accurate probabilistic characteristics of the ADC can be achieved by the proposed methodology with much less computational efforts compared to the LHS-based MCS. 
\end{itemize}

The obtained ADC provides important insights regarding how the uncertainty affects the delivery capability of the distribution network. The simulation results show that the ADC has to be decreased to prevent the system from unexpected violations due to the randomness in the system. Considering the general trend of increasing penetration of RES, it is of paramount significance to carefully consider the resulting increasing variability in a system and to seriously investigate their impact to the performance of a power grid from different perspectives. 

It should be noted that there is a similar concept ``hosting capability (HC)'' emerging in recent publications (e.g., \cite{Bollen11}, \cite{Etherden14}), which is defined as ``the amount of new production or consumption that can be connected to the grid without endangering the reliability or voltage quality for other customers''.
Various performance indices have been embedded into HC assessment, such as voltage quality \cite{Bollen11}, overloading \cite{Haesen05}, and harmonic distortion \cite{Bollen06}, etc. However, there is a fundamental difference between the hosting capacity and the available delivery capability considered in this paper. The hosting capacity typically works as planning tool, which applies repetitive power flow study to identify a relatively conservative amount of distributed generator (DG) deployment from randomly selected combinations of DG locations and sizes. 
In contrast, the ADC discussed in this paper serves as an operation tool, which adopts continuation power flow study to determine the maximum power margin without the violation of operational limits and should be assessed constantly on hour basis.  
For instance, given the forecast of load increasing pattern, along with the probabilistic models and parameter values of the renewable generators in the next two hours, the estimated available delivery capability is able to determine whether the distribution system is able to support such amount of power increase with high probability, and how far the system is away from the thermal limits, the voltage limits and the voltage collapse point. 

The rest of the paper is organized as the following. Section II introduces the probabilistic models of wind power, solar PV and loads. This is followed by the mathematical formulation of probabilistic ADC in Section III. Section IV describes the sparse PCE method and its implementation in assessing ADC. The detailed algorithm to assess the probabilistic ADC is presented in Section V. The simulation results on the IEEE 13 node and 123 node test feeders are given in Section VI. Conclusions and perspectives are discussed in Section VII. 



\section{The Probabilistic Power System Model}

Typically, each of uncertain model input can be expressed as a random variable associated with a probabilistic density function (PDF) ${X\sim f_{X}(x)}$.  
In the context of power grids, the uncertainties mainly come from wind generation, solar generation and load demand. The corresponding random variables are wind speed ${v}$, solar radiation ${r}$ and load active power ${P_{L}}$. In this section, we describe their probabilistic model based on different time ranges of interests. 

\subsection{Wind Generation}
In long-time scale, the Weibull distribution  is a good fit to the observed wind speed empirical distribution in many locations around the world \cite{Karki06}-\cite{Xiaozhe17}. The probability density function for a Weibull distribution is as follows:
\begin{small}
$ f_{V}\left( v \right)=\frac{k}{c}{{\left( \frac{v}{c} \right)}^{k-1}} \exp \left[ -{{\left( \frac{v}{c} \right)}^{k}} \right] $
\end{small}
where $v$ is the wind speed, $k$ and $c$ are the shape and scale parameters, respectively.

In short-time operation analysis, however, the wind speed follows normal distribution \cite{Ran16}:
\begin{small}
$ f_{V}(v)=\frac{1}{\sqrt{2\pi}{{\sigma}_{v}}} \exp \left( -\frac{{{(v-{\mu}_{v})}^{2}}}{2{\sigma}_{v}^{2}} \right) $
\end{small}
where the mean ${{\mu}_{v}}$ is the forecasted wind speed, and variance ${{\sigma}_{v}}$ represents the forecasting error which can be obtained from historical data.

For each realization of wind speed $v$, the corresponding wind power injected into its terminal bus can be calculated by the wind speed-power output relation \cite{Roy02}: 
\begin{small}
\begin{equation}
\label{eq:wind_p_val}
{{P}_{w}}(v)=\left\{ \begin{array}{*{35}{l}}
0 & v\le {{v}_{in}} \mbox{ or } v>{{v}_{out}} \\
\frac{v-{{v}_{in}}}{{{v}_{rated}}-{{v}_{in}}}{{P}_{r}} & {{v}_{in}}<v\le {{v}_{rated}} \\
{{P}_{r}} & {{v}_{rated}}<v\le {{v}_{out}} \\
\end{array} \right.
\end{equation}
\end{small}

\noindent where ${{v}_{in}}$, ${{v}_{out}}$ and ${{v}_{rated}}$ are the cut-in, cut-out, and rated wind speed ($m/s$), ${{P}_{r}}$ is the rated wind power ($kW$) and ${{P}_{w}}$ is the active power output of wind power. The reactive power output ${{Q}_{w}}$ can be computed under the assumption of constant power factor (e.g., 0.85 lagging).

\subsection{Solar Generation}
For long-time scale analysis, the solar radiation is typically represented by Beta distribution \cite{Salameh95}:
\begin{small}
$ f_{R}\left( r \right)=\frac{\Gamma(\alpha+\beta)}{\Gamma(\alpha)\Gamma(\beta)}{{\left( \frac{r}{{{r}_{\max}}} \right)}^{\alpha-1}}{{\left( 1-\frac{r}{{{r}_{\max}}} \right)}^{\beta-1}} $
\end{small}
where $r$ and ${{r}_{\max}}$ (${W/m}^{2}$) are the actual and maximum solar radiations, respectively; $\alpha$ and $\beta$ are the shape parameters of the distribution; $\Gamma$ is the Gamma function. 

For short-time scale application, similar to the wind generation, the solar radiation follows the normal distribution assuming that the mean of solar radiation can be accurately forecasted \cite{Ran16}.
The solar radiation-power output relation is described by \cite{Marwali98}:
\begin{small}
\begin{equation}
\label{eq:solar_p_val}
{{P}_{pv}}(r)=\left\{ \begin{array}{*{35}{l}}
\frac{{{r}^{2}}}{{{r}_{c}}{{r}_{std}}}{{P}_{r}} & 0\le r<{{r}_{c}}  \\
\frac{r}{{{r}_{std}}}{{P}_{r}} & {{r}_{c}}<r\le {{r}_{std}}  \\
{{P}_{r}} & r>{{r}_{std}}  \\
\end{array} \right. 
\end{equation}
\end{small}

\noindent where ${{r}_{c}}$ is a certain radiation point set usually as 150 $W/m^2$,  ${{r}_{std}}$ is the solar radiation in the standard environment, ${{P}_{r}}$ is the rated capacity of the solar PV. Solar generation is injected into the power grid at unity power factor \cite{WECC10}, and hence ${{Q}_{pv}}$ is assumed to be zero in this study.

\subsection{Load Variation}

Rapid deployment of smart meters provides massive amount of real-time residential load data to utility companies, which typically in 15 minutes, 30 minutes or 1 hour resolutions \cite{Cooper16}. Such abundance of data makes short-term load forecasting possible in distribution systems \cite{Ghofrani11}, \cite{Li16}, \cite{Khan17}. To describe the uncertainty of load forecast, the most common practice is to use the normal distribution \cite{Billinton08}. The forecasted mean value ${{\mu}_{P_{L}}}$ of ${{P}_{L}}$ is typically provided by load forecaster, and ${{\sigma}_{P_{L}}}$ denotes the forecasting error. Generally, the load forecaster only provides the active power, whereas the reactive power is determined under the assumption of constant power factor.

It is worth mentioning that the increasing prevalence of RES, especially residential solar PVs, and the varying tariffs structure have noteworthy impact on the load consumption patterns \cite{Caves84}. Users tend to shift usage and respond to the price to save cost, making the load patterns more prone to change than before. However, as distribution systems become fully deployed with smart meters, it is still possible, with a reasonable accuracy, to obtain probability distributions within a short period of time, e.g. a few hours.

%
\section{Mathematical Formulation of Probabilistic Available Delivery Capability}
The three-phase distribution power flow equations can be represented as
\begin{small}
$ {\bm{f}}\left( \bm{x} \right)=\begin{bmatrix}
{P_{i0}^{\varphi}-P_{i}^{\varphi}(\bm{x})} \\ 
{Q_{i0}^{\varphi}-Q_{i}^{\varphi}(\bm{x})} \end{bmatrix}=0 $
\end{small}
where ${\bm{x}}={{\left[ {{\theta}_{a}},{{\theta}_{b}},{{\theta}_{c}},{{V}_{a}},{{V}_{b}},{{V}_{c}} \right]}^{T}}$, e.g., voltage angles and magnitudes for all phases.
Let $v$, $r$ and ${{P}_{L}}$ be the random vectors that represent wind speeds, solar radiations and load variations, respectively, the three phase probabilistic continuation power flow (PCPF) equations of a $N$ bus system can be described as below. Specifically, for PQ type nodes, the PCPF equations are:
\begin{small}
\begin{equation}
\label{eq:pq_bus_p}
\begin{aligned}
P_{i0}^{\varphi} & - V_{i}^{\varphi}\sum\limits_{j=1}^{N}{\sum\limits_{k=1}^{M}{V_{j}^{k}\left(G_{ij}^{\varphi k}\cos \theta_{ij}^{\varphi k}+B_{ij}^{\varphi k}\sin \theta_{ij}^{\varphi k} \right)}} \\
& + \lambda \left(\Delta P_{Gi}^{\varphi}+\Delta P_{wi}^{\varphi}({{v}_{i}})+\Delta P_{pvi}^{\varphi}({{r}_{i}})-\Delta P_{Li}^{\varphi}(P_{Li}^{\varphi})\right)=0
\end{aligned} 
\end{equation}
\end{small}
\begin{small}
\begin{equation}
\label{eq:pq_bus_q}
\begin{aligned}
Q_{i0}^{\varphi} & - V_{i}^{\varphi}\sum\limits_{j=1}^{N}{\sum\limits_{k=1}^{M}{V_{j}^{k}\left(G_{ij}^{\varphi k}\sin \theta_{ij}^{\varphi k}-B_{ij}^{\varphi k}\cos \theta_{ij}^{\varphi k} \right)}} \\ 
& + \lambda \left(\Delta Q_{Gi}^{\varphi}+\Delta Q_{wi}^{\varphi}({{v}_{i}})-\Delta Q_{Li}^{\varphi}(P_{Li}^{\varphi}) \right)=0 
\end{aligned} 
\end{equation}
\end{small}
\noindent For PV type nodes, the corresponding PCPF equations are:
\begin{small}
\begin{equation}
\label{eq:pv_bus_p}
\begin{aligned}
P_{i0}^{\varphi} & - V_{i}^{\varphi}\sum\limits_{j=1}^{N}{\sum\limits_{k=1}^{M}{V_{j}^{k}\left(G_{ij}^{\varphi k}\cos \theta _{ij}^{\varphi k}+B_{ij}^{\varphi k}\sin \theta _{ij}^{\varphi k} \right)}} \\
& + \lambda \left(\Delta P_{Gi}^{\varphi}+\Delta P_{wi}^{\varphi}({{v}_{i}})+\Delta P_{pvi}^{\varphi}({{r}_{i}})-\Delta P_{Li}^{\varphi}(P_{Li}^{\varphi})\right)=0
\end{aligned} 
\end{equation}
\end{small}
\begin{small}
\begin{equation}
\label{eq:pv_bus_v}
V_{i}^{\varphi}={{V}_{i0}} 
\end{equation}
\end{small}
\begin{small}
\begin{equation}
\label{eq:pv_bus_q}
\begin{aligned}
Q_{Gi}^{\varphi} & - Q_{Li0}^{\varphi}-V_{i}^{\varphi}\sum\limits_{j=1}^{N}{\sum\limits_{k=1}^{M}{V_{j}^{k}\left(G_{ij}^{\varphi k}\sin \theta_{ij}^{\varphi k}-B_{ij}^{\varphi k}\cos \theta_{ij}^{\varphi k} \right)}} \\ 
& + Q_{wi0}^{\varphi} + \lambda \left(\Delta Q_{wi}^{\varphi}({{v}_{i}})-\Delta Q_{Li}^{\varphi}(P_{Li}^{\varphi}) \right)=0
\end{aligned} 
\end{equation}
\end{small}
\begin{small}
\begin{equation}
\label{eq:pv_bus_qlimit}
{{Q}_{min,i}}\le Q_{Gi}^{\varphi}\le {{Q}_{max,i}} 
\end{equation}
\end{small}
\noindent where $G_{ij}^{\varphi k}$ and $B_{ij}^{\varphi k}$ are entries in the bus admittance matrix; $\Delta P_{wi}^{\varphi}({{v}_{i}})$, $\Delta P_{pvi}^{\varphi}({{r}_{i}})$, $\Delta P_{Li}^{\varphi}$ and $\Delta P_{Gi}^{\varphi}$ are the real power variation from wind power, solar PV, load and other types of distributed generators at the phase $\varphi$ of bus $i$ respectively; $\Delta Q_{wi}^{\varphi}({{v}_{i}})$ and $\Delta Q_{Li}^{\varphi}$ are the reactive power variation from wind power and load, respectively; $Q_{Gi}^{\varphi }$ is the reactive generation, and $M$ is the number of phases. If $Q_{Gi}^{\varphi}$ exceeds its limits, say ${{Q}_{min,i}}$ or ${{Q}_{max,i}}$, then the terminal bus switches from PV to PQ with $Q_{Gi}^{\varphi}$ fixed at the violated limit.

In fact, the set of parameterized three-phase PCPF equations (\ref{eq:pq_bus_p})-(\ref{eq:pv_bus_qlimit}) can be described in the following compact form:
\begin{small}
\begin{equation}
\label{eq:cpf_equation}
{\bm{f}} \left( {\bm{x},\bm{\mu},\bm{\lambda},\bm{U}} \right)={\bm{f}}\left( {\bm{x},\bm{\mu}} \right)-\lambda {\bm{b}}({\bm{U}})=0 
\end{equation}
\end{small}
\noindent where ${\bm{x}}$ is the state vector, ${\bm{\mu}}$ is the control parameters vector such as the tap ratio of transformers, ${\bm{U}=\left[ \bm{v},\bm{r},\bm{{P}_{L}} \right]}$ is the random vector describing the wind speed, the solar radiation, the load active power. Besides, the load-generation variation vector ${\bm{b}}$ of the system is
\begin{small}
\begin{equation}
\label{eq:cpf_variation}
{\bm{b}} \left( {\bm{U}} \right)=\begin{bmatrix}
\Delta {\bm{P}}_{G}^{\varphi}+\Delta {\bm{P}}_{w}^{\varphi}(v)+\Delta {\bm{P}}_{pv}^{\varphi}(r)-\Delta {\bm{P}}_{L}^{\varphi}({{P}_{L}})  \\
\Delta {\bm{Q}}_{G}^{\varphi}+\Delta {\bm{Q}}_{w}^{\varphi}(v)-\Delta {\bm{Q}}_{L}^{\varphi}({{P}_{L}})  \\
\end{bmatrix} 
\end{equation}
\end{small}
It is obvious that the set of the parameterized power flow equations become the base-case power flow equation if $\lambda =0$.

The probabilistic ADC formulation therefore can be proposed as the following:
\begin{small}
\begin{equation}
\label{eq:prob_adc}
\begin{aligned}
& \text{max } \lambda & & \\
& \text{s.t.} & & {\bm{f}}\left( {\bm{x}},{\bm{\mu}} \right)-\lambda {\bm{b}}({\bm{U}})=0 & (a) \\
& & & {{V}_{min}}\le V_{i}^{\varphi}\left( {\bm{x},\bm{\mu},\bm{\lambda},\bm{U}} \right)\le {{V}_{max}} & (b) \\
& & & I_{ij}^{\varphi}\left( {\bm{x},\bm{\mu},\bm{\lambda},\bm{U}} \right)\le {{I}_{ij,max}} & (c) \\
& & & {{Q}_{min,i}}\le Q_{Gi}^{\varphi }\left( {\bm{x},\bm{\mu},\bm{\lambda},\bm{U}} \right)\le {{Q}_{max,i}} & (d)
\end{aligned} 
\end{equation}
\end{small}

\noindent where ${{V}_{min}}$ and ${{V}_{max}}$ are the lower and upper limits of bus voltages; ${{I}_{ij,max}}$ is the specified capacity of the line or transformer between bus $i$ and bus $j$; ${\lambda}$ is the normalized load margin under given load-generation variation vector. The maximum value of $\lambda$ that could be achieved without the violation of (\ref{eq:prob_adc}) corresponds to the ADC.  Note that ${\lambda}$ is a random variable due to the random input ${\bm{U}}$. Equation (a) specifies that the solution must satisfy the parameterized power flow equations (\ref{eq:cpf_equation}); Equations (b)-(d) imply that the solution has to satisfy typical operational and electrical constrains.

%
\section{Adaptive Sparse Polynomial Chaos Expansion}

In this paper, we will improve and apply the adaptive sparse PC expansion to solve the probabilistic ADC problem.  
The PC expansion and its associated advanced techniques will be detailed in this section, which can handle practical problems with a large number of dependent random inputs accurately, efficiently and reliably. 

\subsection{Generalized Polynomial Chaos Expansion}
Consider a random vector ${\bm{\xi}}$ = (${{\xi}_{1},{\xi}_{2},...,{\xi}_{n},}$) with $n$ independent components and described by the joint probability density function (PDF) ${f_{\bm{\xi}}}$ ($\bm{\xi}$ is related with $\bm{U}$ in (\ref{eq:prob_adc})), then the target stochastic response ${Y}$ (e.g., the ADC in this study) can be represented by:
\begin{small}
\begin{equation}
\label{eq:gpce_full}
Y=g(\bm{\xi})=\sum\limits_{k=0}^{\infty}c_{k}{\Psi }_{k}(\bm{\xi}) 
\end{equation}
\end{small}

\noindent which is termed as the polynomial chaos expansion, where $\{{\bm{\Psi}}_k\}$ is multivariate polynomials orthogonal to $f_{\bm{\xi}}$ and can be constructed as the tensor product of their univariate counterparts \cite{Blatman09}:
\begin{small}
\begin{equation}
\label{eq:gpce_multipoly}
{\bm{\Psi}_{k}}(\bm{\xi})=\prod\limits_{i=1}^{n}{\phi_{{i}_{j}}({{\xi}_{i}})}, k=0,...,\infty 
\end{equation}
\end{small}

\noindent where the subscript ${i_{j}}$ refers to the $j$-th degree of the $i$-th univariate polynomial basis. $\{\phi_{i_j}\}$ is the orthogonal basis with respect to $f_{\xi_i}$. In this notation, there is an implicit one-to-one mapping between the subscript ${k}$ and a multi-index ${a_{k}={(a_{k1},...,a_{kn})}}$ in which each component ${a_{ki}}$ defines the degree $j$ of the $i$-th univariate polynomial basis with ${j}={a_{ki}}$.

Note that for practical implementation, the gPCE of the stochastic response is truncated such that the total degree is not higher than $p$, i.e, $\sum_{j=1}^{M} i_j \leq p$. More advanced truncation strategy will be discussed in Section \ref{sectiontruncation}.
The truncated PC expansion therefore can be expressed as:
\begin{small}
\begin{equation}
\label{eq:gpce_trunc}
Y=g(\bm{\xi}) \approx \sum\limits_{k=0}^{P-1}{{{c}_{k}}{{\Psi }_{k}}(\bm{\xi})}
\end{equation}
\end{small}

\noindent where $P$ is the number of retained terms.

The choice of the type of polynomial ${\phi}_{i}$ depends on the distribution type of the $i$-th random input ${\xi}_{i}$. 
Table \ref{tab:gpce_mapping} shows a set of typical continuous distributions and their corresponding orthogonal polynomials which are able to achieve optimal (exponential) convergence rate \cite{Xiu02}. Numerical study in \cite{Xiu02} has also shown that improper selection of the type of polynomials will lead to clearly slower convergence rate, and result in an higher degree of expansion with much more terms to reach a specified accuracy.
In case a distribution type that is out of the basic types in Table \ref{tab:gpce_mapping}, there are usually two options available. If the distribution of a random input is close to one of basic distributions in Table \ref{tab:gpce_mapping}, then the isoprobabilistic transformation \cite{Lebrun09b} can be applied to projected the random input to a basic distribution so its corresponding polynomials still applicable; otherwise, we could numerically construct a set of univariate polynomial basis in the form:  ${\tilde{\pi}=\frac{{{\pi}_{k}}}{\left\langle {{\pi}_{k}},{{\pi}_{k}} \right\rangle}}$ orthogonal to an arbitrary  distribution. Particularly, the discretized Stieltjes procedure is adopted in this paper which employs a recurrence relation that holds for the computation of any orthogonal polynomial \cite{UQLabPCE17}:

\begin{table}[t]
\renewcommand{\arraystretch}{1.3}
\caption{Standard Forms of Classical Continuous Distributions and their Corresponding Orthogonal Polynomials \cite{Xiu02}}
\label{tab:gpce_mapping}
\begin{center}
\begin{tabular}{c|c|>{\centering}p{1.5cm}|c}
\hline
\bfseries Distribution & \bfseries Density Function & \bfseries Polynomial & \bfseries Support \\
\hline
Normal & ${\frac{1}{\sqrt{2\pi}}{{e}^{-{{x}^{2}}/2}}}$ & Hermite & (-$\infty$,$\infty$) \\
\hline
Uniform & ${\frac{1}{2}}$ & Legendre & [-1,1] \\
\hline
Beta & ${\frac{{{(1-x)}^{\alpha}}{{(1+x)}^{\beta}}}{{{2}^{\alpha+\beta+1}}{B(\alpha+1,\beta+1)}}}$ & Jacobi & [-1,1] \\
\hline
Exponential & ${{e}^{-x}}$ & Laguerre & (0,$\infty$) \\
\hline
Gamma & ${\frac{{{x}^{\alpha}}{{e}^{-x}}}{\Gamma(\alpha+1)}}$ & Generalized Laguerre & [0,$\infty$) \\
\hline
\end{tabular}
\end{center}
\begin{tablenotes}
\item * The Beta function is defined as $B(p,q)=\frac{\Gamma(p)\Gamma(q)}{\Gamma(p+q)}$.
\end{tablenotes}
\end{table}
\begin{small}
\begin{equation}
\label{eq:arbi_poly_recur}
\begin{gathered}
{\sqrt{{b}_{k+1}}}{{\tilde{\pi}}_{k+1}}\left( {\xi_{i}} \right)=\left( {\xi_{i}}-{{d}_{k}} \right){{\tilde{\pi}}_{k}}\left( {\xi_{i}} \right)-\sqrt{{b}_{k}}{{\tilde{\pi}}_{k-1}}\left( {\xi_{i}} \right) \\
{{{{\tilde{\pi}}_{-1}}\left( {\xi_{i}} \right)=0,\tilde{\pi}}_{0}}\left( {\xi_{i}} \right)=1 
\end{gathered}
\end{equation}
\end{small}
\begin{small}
\begin{equation}
\label{eq:arbi_poly_recur_coef}
\begin{gathered}
{{d}_{k}}=\frac{\left\langle \xi {{\pi}_{k}},{{\pi}_{k}} \right\rangle }{\left\langle {{\pi}_{k}},{{\pi}_{k}} \right\rangle }, k \ge 0 \\
{{b}_{0}}=\left\langle {{\pi}_{0}},{{\pi}_{0}} \right\rangle ,{{b}_{k}}=\frac{\left\langle {{\pi}_{k}},{{\pi}_{k}} \right\rangle }{\left\langle {{\pi}_{k-1}},{{\pi}_{k-1}} \right\rangle }, k \ge 1  
\end{gathered}
\end{equation}
\end{small}

\noindent where $k$ denotes the polynomial degree.

To illustrate the construction of mixed multivariate polynomials, consider a 3-dimensional input random vector ${\xi=(\xi_{1},\xi_{2},\xi_{3})}$ with independent components ${{{\xi}_{1}}\sim \mbox{Weibull}\left( \lambda,k \right)}$, ${{{\xi}_{2}}\sim \mbox{Beta}\left( \alpha,\beta \right)}$ and ${{{\xi}_{3}}\sim N\left( 0,1 \right)}$ respectively, the multivariate polynomials that correspond to ${p \le 2}$ are:
\begin{small}
\begin{equation}
\label{eq:multiv_poly_demo}
\begin{aligned}
{{\Psi}_{{0}}}\left( \bm{\xi} \right) &={{{\tilde{\pi}}}_{1_0}}\left( {{\xi}_{1}} \right){{\phi}_{2_0}}\left( {{\xi}_{2}} \right){{\phi}_{3_0}}\left( {{\xi}_{3}} \right) \quad \mapsto \quad a_{0}=(0,0,0) \\ 
{{\Psi}_{{1}}}\left( \bm{\xi} \right) &={{{\tilde{\pi}}}_{1_1}}\left( {{\xi}_{1}} \right){{\phi}_{2_0}}\left( {{\xi}_{2}} \right){{\phi}_{3_0}}\left( {{\xi}_{3}} \right) \quad \mapsto \quad a_{1}=(1,0,0) \\ 
{{\Psi}_{{2}}}\left( \bm{\xi} \right) &={{{\tilde{\pi}}}_{1_0}}\left( {{\xi}_{1}} \right){{\phi}_{2_1}}\left( {{\xi}_{2}} \right){{\phi}_{3_0}}\left( {{\xi}_{3}} \right) \quad \mapsto \quad a_{2}=(0,1,0) \\ 
{{\Psi}_{{3}}}\left( \bm{\xi} \right) &={{{\tilde{\pi}}}_{1_0}}\left( {{\xi}_{1}} \right){{\phi}_{2_0}}\left( {{\xi}_{2}} \right){{\phi}_{3_1}}\left( {{\xi}_{3}} \right) \quad \mapsto \quad a_{3}=(0,0,1) \\ 
{{\Psi}_{{4}}}\left( \bm{\xi} \right) &={{{\tilde{\pi}}}_{1_1}}\left( {{\xi}_{1}} \right){{\phi}_{2_1}}\left( {{\xi}_{2}} \right){{\phi}_{3_0}}\left( {{\xi}_{3}} \right) \quad \mapsto \quad a_{4}=(1,1,0) \\ 
{{\Psi}_{{5}}}\left( \bm{\xi} \right) &={{{\tilde{\pi}}}_{1_1}}\left( {{\xi}_{1}} \right){{\phi}_{2_0}}\left( {{\xi}_{2}} \right){{\phi}_{3_1}}\left( {{\xi}_{3}} \right) \quad \mapsto \quad a_{5}=(1,0,1) \\ 
{{\Psi}_{{6}}}\left( \bm{\xi} \right) &={{{\tilde{\pi}}}_{1_0}}\left( {{\xi}_{1}} \right){{\phi}_{2_1}}\left( {{\xi}_{2}} \right){{\phi}_{3_1}}\left( {{\xi}_{3}} \right) \quad \mapsto \quad a_{6}=(0,1,1) \\ 
{{\Psi}_{{7}}}\left( \bm{\xi} \right) &={{{\tilde{\pi}}}_{1_2}}\left( {{\xi}_{1}} \right){{\phi}_{2_0}}\left( {{\xi}_{2}} \right){{\phi}_{3_0}}\left( {{\xi}_{3}} \right) \quad \mapsto \quad a_{7}=(2,0,0) \\ 
{{\Psi}_{{8}}}\left( \bm{\xi} \right) &={{{\tilde{\pi}}}_{1_0}}\left( {{\xi}_{1}} \right){{\phi}_{2_2}}\left( {{\xi}_{2}} \right){{\phi}_{3_0}}\left( {{\xi}_{3}} \right) \quad \mapsto \quad a_{8}=(0,2,0) \\ 
{{\Psi}_{{9}}}\left( \bm{\xi} \right) &={{{\tilde{\pi}}}_{1_0}}\left( {{\xi}_{1}} \right){{\phi}_{2_0}}\left( {{\xi}_{2}} \right){{\phi}_{3_2}}\left( {{\xi}_{3}} \right) \quad \mapsto \quad a_{9}=(0,0,2) \\ 
\end{aligned} 
\end{equation}
\end{small}

\noindent where ${\tilde{\pi}_{1}}$, ${\phi_{2}}$ and ${\phi_{3}}$ represent the univariate numerical polynomial, the Jacobi polynomial and the Hermite polynomial, respectively, and their subscripts identify their degrees. Now we have ${Y=c_{0}{\Psi}_{0}({\xi})+...+c_{9}{\Psi}_{9}({\xi})}$ with known polynomial basis, the next step is to determine the coefficients $\{c_i:i=0,...,9\}$.

\subsection{Calculation of the Coefficients}
The least-square minimization is employed in this paper to calculate the coefficients because of its efficiency for high-dimensional problems. 
Let $\bm{\xi}_{C} = \{\bm{\xi}^{(1)}, \bm{\xi}^{(2)}, ..., \bm{\xi}^{(M_{C})}\}$ be $M_C$ samples of the random variables $\bm{\xi}$, also termed as the experimental design (ED) , and $\bm{y}_{C} = \{y^{(1)}, y^{(2)}, ..., y^{(M_{C})}\}$ be the corresponding stochastic responses calculated by a deterministic method, then the coefficients can be calculated by solving the following least-square minimization problem:
\begin{small}
\begin{equation}
\small
\label{eq:least_sqr}
\hat{\bm{c}}
=\arg \min \sum\limits_{l=1}^{{{M}_{c}}}{{{\left[ \bm{H} \left( {{\xi }^{\left( l \right)}} \right) {\bm{c}}-{{y}^{\left( l \right)}} \right]}^{2}}}
\end{equation}
\end{small}

\noindent where ${\bm{H}}$ is the so-called design matrix evaluated as:
\begin{small}
\begin{equation}
\label{eq:gpce_h_ed}
\bm{H}=\left[ \begin{matrix}
{\bm{\Psi}_{0}({{\bm{\xi}}^{(1)}})} & {\bm{\Psi}_{1}({\bm{\xi}^{(1)}})} & \cdots & {\bm{\Psi}_{P-1}({\bm{\xi}^{(1)}})} \\
{\bm{\Psi}_{0}({\bm{\xi}^{(2)}})} & {\bm{\Psi}_{1}({\bm{\xi}^{(2)}})} & \cdots & {\bm{\Psi}_{P-1}({\bm{\xi}^{(2)}})} \\
\vdots & \vdots & \vdots & \vdots \\
{\bm{\Psi}_{0}({\bm{\xi}^{(M_{C})}})} & {\bm{\Psi}_{1}({\bm{\xi}^{(M_{C})}})} & \cdots & {\bm{\Psi}_{P-1}({\bm{\xi}^{(M_{C})}})} \\
\end{matrix} \right] 
\end{equation}
\end{small}

\noindent where ${P}$ is the number of retained terms (i.e., ${P=10}$ in (\ref{eq:multiv_poly_demo})).
The least square solution of their corresponding coefficients can be achieved by
\begin{small}
\begin{equation}
\label{eq:solve_least_sqr}
\hat{\bm{c}}=({{\bm{H}}^{T}}{\bm{H}})^{-1}{{\bm{H}}^{T}}{\bm{y}_{C}} 
\end{equation}
\end{small}
\subsection{Least Angle Regression (LAR)-based Sparse Scheme} \label{sectiontruncation}
The standard truncation scheme retains polynomials up to a certain degree ${p}$, however, the number of retained terms ${P}$ increases exponentially with both the number of inputs ${n}$ and the degree ${p}$. 
To mitigate this issue, the maximum interaction and hyperbolic (or ${q}$-norm) truncation \cite{Blatman09} have been employed based on the fact that most of responses depend on main effects and low-rank interactions. 
\begin{small}
\begin{equation}
\label{eq:gpce_q_norm}
{\bm{A}_{n,p,q}}=\{\bm{a}_{k} \in {{N}^{n}}:{{\left( \sum\limits_{i=1}^{n}{a_{ki}^{q}} \right)}^{1/q}}\le p, q \in (0,1)\} 
\end{equation}
\end{small}

In addition, the least angle regression (LAR) is exploited to automatically detect the significant coefficients so that they can be estimated by a small set of simulations, while the rest of coefficients are set to zero. 
This scheme is based on the fact that the predictors (column vector of ${\bm{H}}$ in (\ref{eq:gpce_h_ed})) are not equally relevant in the sense that some predictors may contribute more significantly to the response ${\bm{y}}$ than the others. 
LAR is an effective regression tool for fitting the linear model even when the number of predictors is much larger than the available simulation data (rows of ${\bm{H}}$). The readers are referred to \cite{Efron04} for more details on LAR. Eventually, a sparse PC expansion of the response can be built.
The response (e.g., ADCs in this study) of any new samples can be evaluated efficiently by directly substituting to the solved PC expansion (\ref{eq:gpce_trunc}) instead of solving the original complex problem (e.g., the PCPF problem (\ref{eq:prob_adc})).

It is worthy mentioning that the set of significant coefficients and their estimated values depend on the samples used. Adding new samples may lead to a different set of PCE coefficients. Hence, a sophisticate error index is required to ensure the accuracy and avoid the over-fitting issue as discussed in the next subsection.
\subsection{Adaptive Sparse PCE Procedure}
The aforementioned basis truncation scheme requires an predetermined degree $p$ which is crucial to the accuracy and efficiency, which nevertheless is difficult to choose in advance since it is strongly problem-dependent. Hence, instead of setting a fixed $p$, an adaptive procedure can be employed to a range of candidates ${(p_{0},...,p_{max})}$ to find a satisfactory sparse PC expansion. Given a experimental design of size $M_{C}$, the sparse PC expansion is built with sequentially increasing polynomial degrees, starting from ${p_{0}}$ until the prescribed accuracy or the $p_{max}$ is reached. 

To evaluate the accuracy of the spare PC expansion while avoiding the over-fitting issue, this paper applies the corrected leave-one-out error index $\varepsilon_{cloo}$ as the stopping criteria \cite{Blatman09}:
\begin{small}
\begin{equation}
\label{eq:loo_error_modified}
\varepsilon_{cloo} ={{\varepsilon}_{loo}} \times {T({M_{C}},P)} 
\end{equation}
\end{small}
\noindent where
\begin{small}
\begin{equation}
\label{eq:loo_error}
\begin{gathered}
{{\varepsilon }_{loo}} = \frac{1}{M_{C}} \sum \limits_{i=1}^{M_{C}}{{{\left[ \frac{{y}^{(i)}-{\hat{g}}\left( {\bm{\xi}}^{(i)} \right)}{1-{{h}_{i}}} \right]}^{2}}/\sigma _{Y}^{2}} \\
{T({M_{C}},P)} = \frac{M_{C}}{M_{C}-P} \left\{ 1+tr\left[ {{\left( {{{\bm{H}}}^{T}}{\bm{H}} \right)}^{-1}} \right] \right\}
\end{gathered} 
\end{equation}
\end{small}

\noindent and $\bm{h}=diag\left[ {\bm{H}}{{\left( {{\bm{H}}^{T}}{\bm{H}} \right)}^{-1}}{{\bm{H}}^{T}} \right]$, ${\sigma _{Y}^{2}}$  is the empirical variance of the response vector ${\bm{y}_{C}}$ evaluated on the experimental design.

This error index has two distinguished features: (i) it requires to run model simulation only once to obtain ${{\varepsilon}_{loo}}$ compared to the $n$ times in traditional leave-one-out error evaluation; (ii) it is more effective against the over-fitting issue by introducing a correcting factor ${T({M_{C}},P)}$ proposed in \cite{Chapelle02}.

\subsection{The Nataf Transformation}
So far, the random inputs are assumed to be mutually independent as required by the adaptive sparse PC expansion. To accommodate dependent random inputs, we need to obtain the joint distribution of the variables and build a bridge between the dependent inputs $\bm{U}$ and the independent standard variables $\bm{\xi}$ on which the polynomials basis can be built.

The Nataf transformation offers an effective way to model the dependence structure of a random vector using a normal copula parameterized by its correlation matrix ${\bm{R}}$ \cite{Lebrun09}.
Given the marginal cumulative distribution function ${{F}_{i}}\left( {{u}_{i}} \right)$ of each continuous random variable ${{U}_{i},i=1,2,...,n}$ and the linear correlation matrix ${\bm{\rho}}$, the \textit{Nataf transformation} ${T_{Nataf}(\bm{u})}$ defined by \cite{Lebrun09b}:
\begin{small}
\begin{equation}
\label{eq:Nataf_T_Steps}
\bm{\eta}=T_{Nataf}(\bm{u})=T_{3} \circ T_{2} \circ T_{1}(\bm{u}) 
\end{equation}
\end{small}
\noindent maps $\bm{u}$ to a vector of independent  standard normal random vector $\bm{\eta}$.
Particularly, the three transformations $T_{1}$, $T_{2}$ and $T_{3}$ are:
\begin{small}
\begin{equation}
\label{eq:Nataf_T}
\begin{gathered}
{{T}_{1}}: \bm{u}\mapsto \bm{w}={{\left[ {{F}_{1}}\left( {{u}_{1}} \right),...,{{F}_{n}}\left( {{u}_{n}} \right) \right]}^{T}} \\
{{T}_{2}}: \bm{w}\mapsto \bm{z}={{\left[ {{\Phi }^{-1}}\left( {{w}_{1}} \right),...,{{\Phi }^{-1}}\left( {{w}_{n}} \right) \right]}^{T}} \\ 
{{T}_{3}}: \bm{z}\mapsto \bm{\eta}={\bm{L}^{-1}}{\bm{z}} 
\end{gathered}
\end{equation}
\end{small}

\noindent where ${\Phi}$ is the cumulative distribution function of 1-dimensional normal distribution, ${\bm{L}}$ is the unique lower triangular matrix from Cholesky factorization of ${\bm{R}}$: ${\bm{R}=LL^{T}}$, and ${\bm{R}}$ is the linear correlation matrix of the multivariate normal distribution associated with the normal copula. The off-diagonal element ${R_{ij}}$ can be obtained by solving the following quadrature equations provided that $\rho_{ij}$ is known:
\begin{small}
\begin{equation}
\label{eq:Corr_R_Pearson}
\rho_{ij}= \iint_{{{R}^{2}}}{\left( \frac{{x}_{i}-{\mu}_{i}}{\sigma_{i}} \right)}\left( \frac{{x}_{j}-{\mu}_{j}}{\sigma_{j}} \right){{\Phi }_{2}}\left( {{z}_{i}},{{z}_{j}},{{R}_{ij}} \right)d{{z}_{i}}d{{z}_{j}} 
\end{equation}
\end{small}
\noindent where ${{\mu}_{i}}$ and ${\sigma_{i}}$ are the mean and variance of ${{x}_{i}}$ respectively, ${{{\Phi }_{2}}\left( {{z}_{i}},{{z}_{j}},{{R}_{ij}} \right)}$ is the PDF of a two-dimensional standard normal distribution with mean 0, variance 1, and correlation coefficient ${{R}_{ij}}$. The readers are referred to a novel method in \cite{Shuang08} for solving (\ref{eq:Corr_R_Pearson}).

The Nataf transformation maps the random input ${\bm{U}}$ into an independent standard normal random vector ${\bm{\eta}}$. One valuable property of Nataf transformation is invertibility. In fact, it is the inverse Nataf transformation ${\bm{u}}=T_{Nataf}^{-1} (\bm{\eta})$ that is normally used in experimental design, by which the set of samples of ${\bm{\eta}} $ can be transformed back into samples of ${\bm{U}}$ so that the corresponding responses ${\bm{y}}$ can be evaluated. Note that a further isoprobabilistic transformation \cite{Lebrun09b} may be required if ${\xi_{i}}$ is not a standard normal random variable:
\begin{small}
\begin{equation}
\label{eq:Isoprob_Transf}
T_{4}: {\xi_{i}}={G^{-1}(\Phi(\eta_{i}))} 
\end{equation}
\end{small}
\noindent where $G^{-1}$ is the inverse cumulative distribution function of ${\xi_{i}}$; otherwise $\xi_i=\eta_i$.

It should be pointed out that the presented method in this section is still applicable when $Y$ is a stochastic response vector. In that case, the adaptive sparse PCE procedure will be performed repeatedly to find a satisfactory PC expansion for each of component of $Y$. In fact, there are three responses in this study, i.e., the voltage-violation ADC, thermal-violation ADC and voltage-collapse ADC.


\section{Computation of Probabilistic ADC}

In this section, a step-by-step description of the proposed probabilistic ADC calculation is summarized below:

\noindent Step 1. Input network data, probability distribution and parameters of ${n}$ random inputs, i.e., wind speed, solar radiation and load, and their correlation matrix ${\bm{\rho}}$.

\noindent Step 2. Choose the independent standard variable ${\xi_{i}}$ and corresponding univariate polynomial ${\phi_{i}}$ for each of random input $u_{i}$.

\noindent Step 3. Generate an experimental design of size ${{M}_{C}}$. This process can be implemented as the following steps
\begin{itemize}
\item a) Generation ${M_{C}}$ samples ${\bm{\xi}_{C}=(\xi^{(1)},\xi^{(2)},...,\xi^{({M_{C}})})}$ in standard space by the Latin hypercube sampling (LHS). 
\item b) Transform ${\bm{\xi}_{C}}$ into physical space by the inverse Nataf transformation ${\bm{u}_{C}}=T_{Nataf}^{-1} \circ T_{4}^{-1}(\bm{\xi}_{C})$;
\item c) Evaluate ${\bm{u}_{C}}$ by deterministic simulation tool, say CDFLOW tool \cite{Sheng14b}, to obtain the accurate responses ${\bm{y}_{C}=(y^{(1)},y^{(2)},...,y^{({M_{C}})})}$;
\end{itemize}

The set of sample-response pairs $({\bm{\xi}_{C}},{\bm{y}_{C}})$ can then be used to build the PC expansions of (\ref{eq:gpce_trunc}).

\noindent Step 4. Apply the adaptive sparse procedure to find the best PC expansion for each response ${Y_{i}}$ sequentially.
The $k$th ($p_{0}<k\le p_{max}$) PC expansion of the response $Y_{i}$ can be built by the following steps:
\begin{itemize}
\item a) Generate the set of candidate multi-indices of degree ${k}$ and truncate them using (\ref{eq:gpce_q_norm}), and then evaluate the resulting multivariate polynomials of degree ${k}$ to form the design matrix ${\bm{H}_{k}=[\bm{H}_{k-1},\Delta \bm{H}_{k}]}$.
\item b) Apply the LAR algorithm to select the best set of retained columns of ${\bm{H}_{k}}$, then estimate the corresponding coefficients by (\ref{eq:least_sqr}).
\item c) Calculate the correlated leave-one-out error $e_{cloo}^{(k)}$ by (\ref{eq:loo_error_modified}). If it reaches the prescribed accuracy, i.e., ${e_{cloo}^{(k)} < \epsilon}$ or $e_{cloo}^{(k)} \ge e_{cloo}^{(k-1)} \ge e_{cloo}^{(k-2)}$, then return the best PC expansion. Otherwise, increase the degree of polynomials and go to a).
\end{itemize}

\noindent Step 5. If the best PC expansions of all responses have reached the required accuracy, go to Step 6; otherwise, go to Step 3 to generate additional ${\Delta M_{C}}$ new samples, then go back to Step 4 using the enriched experiment design $({\bm{\xi}_{C}}+{\Delta \bm{\xi}_{C}},{\bm{y}_{C}}+{\Delta \bm{y}_{C}})$. Note that the previous solved PC expansion for each response can be reused as the starting point.

\noindent Step 6. Once the PC expansions for all responses are obtained, sample ${\bm{\xi}}$ extensively, e.g., ${{M}_{S}}$ samples, and apply the solved PC expansions (\ref{eq:gpce_trunc}) to evaluate the corresponding responses ${\bm{Y}}$ for all these samples.

\noindent Step 7. Compute the statistics of each response, and generate the result report.

\noindent Remark: the number of samples ${{M}_{C}}$ in Step 3 is usually much smaller than ${{M}_{S}}$ in Step 6. Unlike MCS, PCE does not solve (\ref{eq:prob_adc}) for the ${{M}_{S}}$ samples in Step 6, hence it is more efficient. The main computational effort of PCE lies in Step 3.


\section{Numerical Studies}

In this section, we apply the proposed method to investigate the probabilistic ADC of the modified IEEE 13 node and 123 node test feeders \cite{IEEE92}.
The LHS-based Monte Carlos simulation (MCS) is used as a benchmark to validate the accuracy and the performance of the proposed method.
In this study, we assume that wind speed follows Weibull distribution, solar radiation follows Beta distribution, and load power follows normal distribution, the parameters of which are available on the study period. Particularly, the shape and scale parameters of the Weibull distribution for wind speed are $k=7.41$ and $c=2.06$; the shape parameters and boundaries of the Beta distribution are $\alpha=2.06$, $\beta=2.50$, $r=0$ and $s=1000$. For each load, the mean of load increase is set to the base case value and the variance equals to 5\% of its mean.
The corresponding univariate polynomials of Weibull distribution, Beta distribution and normal distribution are chosen to be numerical polynomial, Jacobi polynomial and Hermite polynomial, respectively. It should be noted that the proposed method is not limited by any specific marginal probability distribution.

For simplicity, all wind generators share the same set of $v_{rate}$, $v_{in}$, and $v_{out}$ which are 15.0, 4.0, and 25.0 ${m/s}$, respectively. Similarly, all solar PVs are assumed to share the same $r_{c}$ and $r_{std}$, which are 150.0 and 1000.0 ${W/m^{2}}$, respectively. 
The linear correlation coefficient ${\rho_{ij}}$ between component $i$ and $j$ of wind speed $v$, solar radiations $r$ and load power $P_{L}$ are 0.5, 0.8, 0.4, respectively.

\subsection{The Modified IEEE 13 Node Test Feeder}
We add two solar PVs and two wind generators to the IEEE 13 node test feeder, the total loads of which are 1.733 MW and 1.051 Mvar. The rated power $P_{r}$ of the solar PVs at bus 675 and bus 692 are 180kW and 240kW, respectively, and the rated power $P_{r}$ of the wind generators at bus 680 and bus 634 are 450kW and 300kW, respectively. 
Since there are 8 single-phase loads in this feeder, the total random inputs is 12. 

\begin{figure}[h]
\centering
\includegraphics[width=0.35\textwidth]{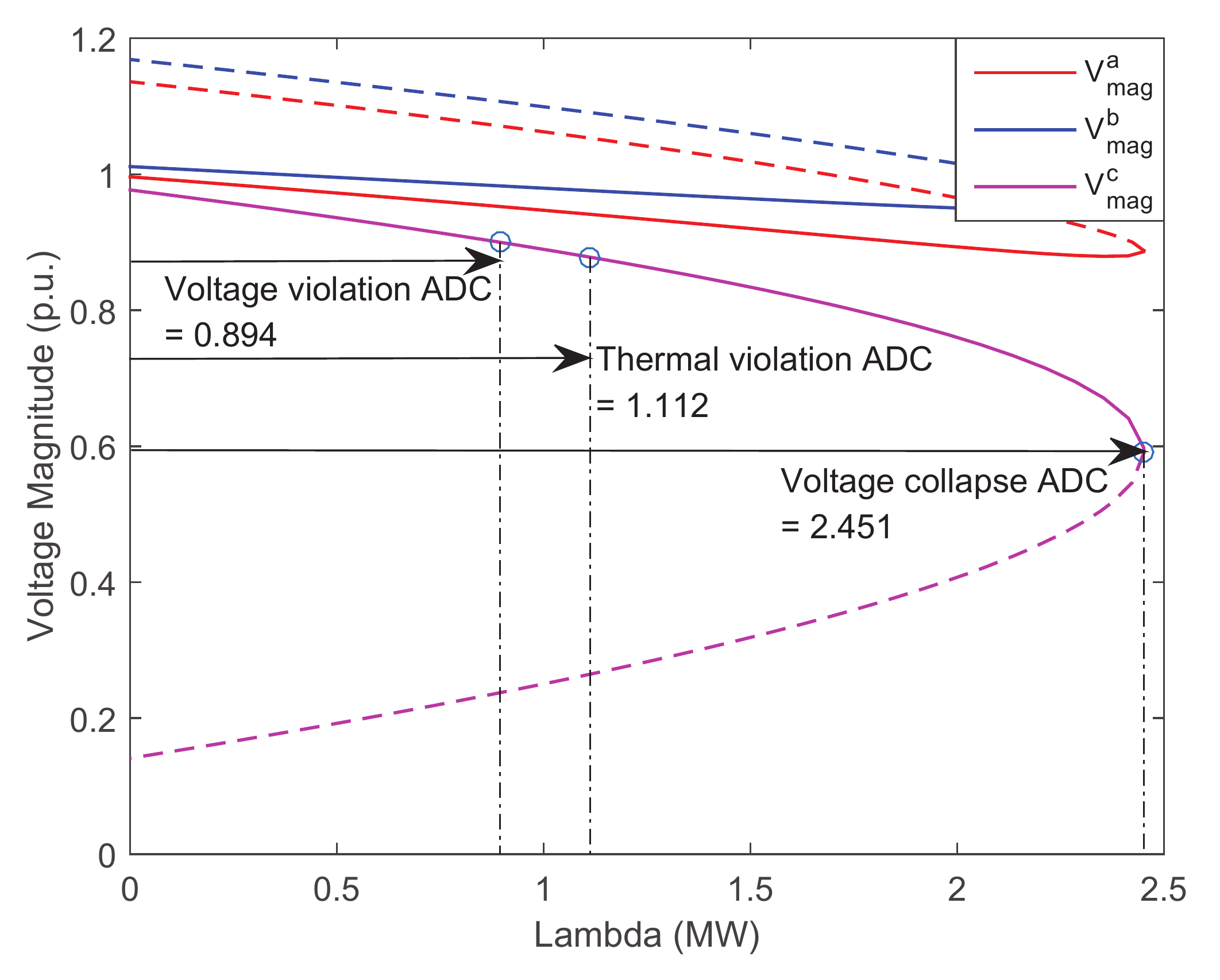}
\caption{Without considering uncertainties, the voltage-violation ADC is 0.894 MW, the thermal-violation ADC is 1.112 MW and the voltage-collapse ADC is 2.451 MW. The (overall) ADC is hence 0.894 MW.}
\label{fig:adc_base}
\end{figure}

To investigate the impact of uncertainty on the ADC of the system, we first apply the CDFLOW tool to trace the P-V curves and ADCs of the deterministic system, i.e., without uncertainty. 
Fig. \ref{fig:adc_base} shows the three-phase P-V curves of bus 675 at the sample point where all random inputs equal to their mean values (i.e., deterministic system). The P-V curves of different phases are quite different due to the unbalanced network parameters and loads. 
Voltage magnitude at phase C drops the most quickly as the load-generation increase along the predetermined direction. 
Among the three ADCs (the ADC subject to voltage violation, the ADC subject to thermal violation, and the ADC subject to voltage collapse), the ADC subject to voltage violation, 0.894 MW, is the smallest one due to fact that the voltage at the single-phase bus 611 reaches the lower voltage limit 0.90 p.u. Hence the overall ADC of the system is 0.894 MW.

Next, we assess the probabilistic ADC by the proposed method and compare the results with the benchmark MCS.
Based on the experiment design, the LAR algorithm truncates the terms of the full PC expansion \eqref{eq:gpce_full} after the second degree, leading to 25 multivariate polynomials, i.e., $P=25$ in (\ref{eq:gpce_trunc}).
Then the least square method is applied to estimate the coefficients corresponding to the retained multivariate polynomials. 
After the coefficients of sparse PC expansion (\ref{eq:gpce_trunc}) are solved, 4000 samples are generated to assess the statistical properties of ADCs by the MCS and the solved sparse PC expansion, respectively.
Table \ref{tab:adc_comp} shows the estimated mean and variance of the ADCs by the MCS and the sparse PCE, from which we can see that the sparse PCE is able to provide reasonably good estimations.

\begin{table}[]
\renewcommand{\arraystretch}{1.3}
\caption{Comparison of the Estimated Statistics of the ADC}
\label{tab:adc_comp}
\centering
\begin{tabular}{c|c|c|c||c|c|c}
\hline
ADC & $\mu_{mcs}$ & $\mu_{pce}$ & ${\frac{\Delta \mu}{\mu_{mcs}}} \%$ & $\sigma_{mcs}^{2}$ & $\sigma_{pce}^{2}$ & ${\frac{\Delta \sigma^{2}}{\mu_{mcs}}} \%$ \\
%
%
\hline
V.V. & 0.8935 & 0.8935 &  0.0023 & 0.0003 & 0.0002 & -0.0013  \\
\hline
T.V. & 1.1143 & 1.1138 & -0.0440 & 0.0039 & 0.0038 & -0.0067  \\
\hline
V.C. & 2.4521 & 2.4514 & -0.0312 & 0.0033 & 0.0023 & -0.0407  \\
\hline
\end{tabular}
\begin{tablenotes}
\item * The V.V., T.V. and V.C. represent voltage violation, thermal violation and voltage collapse respectively.
\end{tablenotes}
\end{table}

Nevertheless, to get these comparable accuracy, MCS needs to run 4000 simulations (i.e. solving (\ref{eq:prob_adc})), while the sparse PCE only requires 31 simulations for solving coefficients of (\ref{eq:gpce_trunc}). Table \ref{tab:time_comp} lists the time for experimental design ${t_{ed}}$, solving coefficients ${t_{sc}}$, and evaluating statistic samples ${t_{es}}$, and the total time ${t_{total}}$. The time that sparse PCE spends on solving coefficients and evaluating statistic samples is negligible compared to that for solving (\ref{eq:prob_adc}) for 4000 times. Clearly, the sparse PCE is much more efficient than MCS.
\begin{table}[]
\renewcommand{\arraystretch}{1.3}
\caption{Comparison of Computation Time between SPCE and MCS}
\label{tab:time_comp}
\centering
\begin{tabular}{c|c|c|c|c}
\hline
Method & ${t_{ed}(s)}$ & ${t_{sc}(s)}$ & ${t_{es}(s)}$ & ${t_{total}(s)}$ \\
\hline
SPCE   & 8.713  &  0.934  &  0.213   &  9.860 \\
\hline
MCS    & --     & --      & 2810.654 & 2810.654 \\
\hline
\end{tabular}
\end{table}
\begin{figure}[h]
\centering
\includegraphics[width=0.45\textwidth]{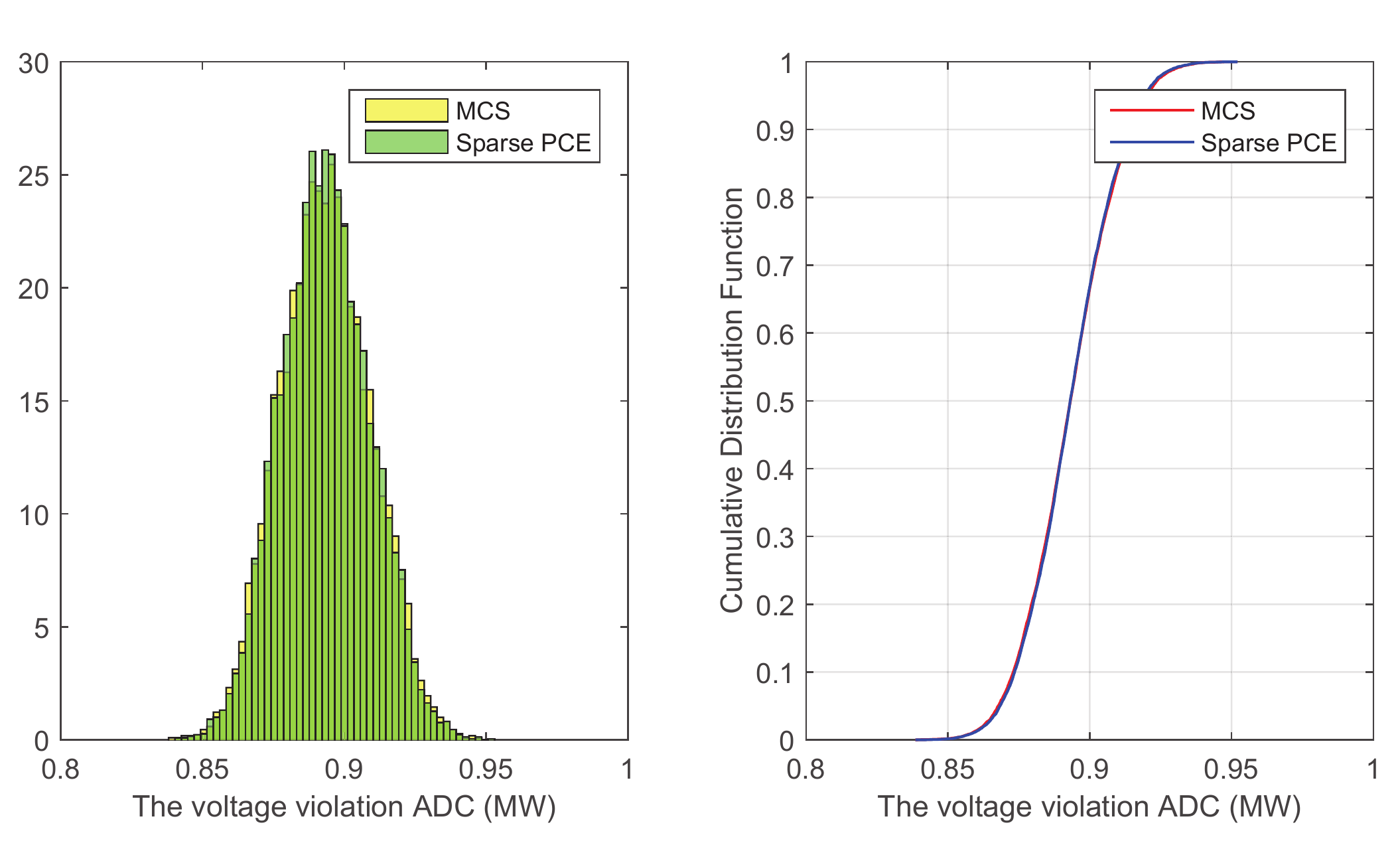}
\caption{The distribution of voltage-violation ADC computed by MCS and sparse PCE. They are almost overlapped. To ensure 95\% confidence level that the system will not encounter unexpected voltage violations, the ADC has to be reduced 2.91\% of the value calculated in deterministic system.}
\label{fig:adc_dist}
\end{figure}

Once we have the statistics of the probabilistic ADC, one natural step afterwards is to see how the uncertainty actually affect the delivery capability of a generic distribution system. As shown in Fig. \ref{fig:adc_base}, the voltage-violation ADC without considering the uncertainty is 0.894MW, it is interesting to see that the corresponding probability at the CDF curve (Fig. \ref{fig:adc_dist}) is 0.518, indicating that there is 51.8\% probability that the delivery capability is less than or equal to 0.894MW. 
Such low probability of security is typically unacceptable in real-world applications. To ensure 95\% confidence level that the system will not encounter voltage violation, however, the voltage-violation ADC has to be reduced from 0.894MW to 0.868MW corresponding to 2.91\% reduction. These results clearly show that the overall ADC has to be reduced to account for the variability of RES and load variations.

\subsection{The Modified IEEE 123 Node Test Feeder}
The IEEE 123 node test feeder is a comprehensive unbalanced feeder which has overhead and underground lines with various phasing, regulators, shunt capacitor banks, and unbalanced loading with all combination of load types (constant PQ, constant I and constant Z) \cite{IEEE92}. In order to assess the impact of uncertainties on the ADC of this network, 20 solar PVs and 20 wind generator of 30kW are connected to the network. There are totally 134 random inputs including stochastic loads.  

Applying the full second-order PC expansion, we need to run thousands of simulations to calculate the 9180 coefficients, which becomes a huge burden and weakens the advantage of the PCE over the MCS. To mitigate the issue, the sparse PCE is applied, which requires only 336 simulations to solve the 269 coefficients of the truncated PC expansions (\ref{eq:gpce_trunc}). In contrast, MCS requires 10000 simulation (i.e. solving (\ref{eq:prob_adc})) to get comparable accuracy. 
Table \ref{tab:adc_comp_123} presents a comparison between the estimated mean, variance, and percentage error of the probabilistic ADC evaluated by the proposed sparse PCE and those by the MCS. Table \ref{tab:time_comp_123} shows the time spent by the two methods. Clearly, these results demonstrate that the proposed sparse PCE is able to provide accurate estimation for the probabilistic ADC using much less computational time ($\approx \frac{1}{30}$) than the MCS. 
\begin{table}[]
\renewcommand{\arraystretch}{1.3}
\caption{Comparison of the Estimated Statistics of the ADC}
\label{tab:adc_comp_123}
\centering
\begin{tabular}{c|c|c|c||c|c|c}
\hline
ADC & $\mu_{mcs}$ & $\mu_{pce}$ & ${\frac{\Delta \mu}{\mu_{mcs}}} \%$ & $\sigma_{mcs}^{2}$ & $\sigma_{pce}^{2}$ & ${\frac{\Delta \sigma^{2}}{\mu_{mcs}}} \%$ \\
%
%
\hline
V.V. & 19.2794 & 19.2808 &  0.0070 & 0.2002 & 0.1976 & -0.0131 \\
\hline
T.V. & 2.0710  & 2.0716  &  0.0277 & 0.0048 & 0.0043 & -0.0231 \\
\hline
V.C. & 58.4813 & 58.4697 & -0.0199 & 2.9956 & 2.9943 & -0.0021 \\
\hline
\end{tabular}
\end{table}
\begin{table}[]
\renewcommand{\arraystretch}{1.3}
\caption{Comparison of Computation Time between SPCE and MCS}
\label{tab:time_comp_123}
\centering
\begin{tabular}{c|c|c|c|c}
\hline
Method & ${t_{ed}(s)}$ & ${t_{sc}(s)}$ & ${t_{es}(s)}$ & ${t_{total}(s)}$ \\
%
%
\hline
SPCE   & 732.144 & 28.796 &     0.652 &   761.592 \\
\hline
MCS    & --      & --     & 21790.000 & 21790.000 \\
\hline
\end{tabular}
\end{table}
\begin{figure}[h]
\centering
\includegraphics[width=0.45\textwidth]{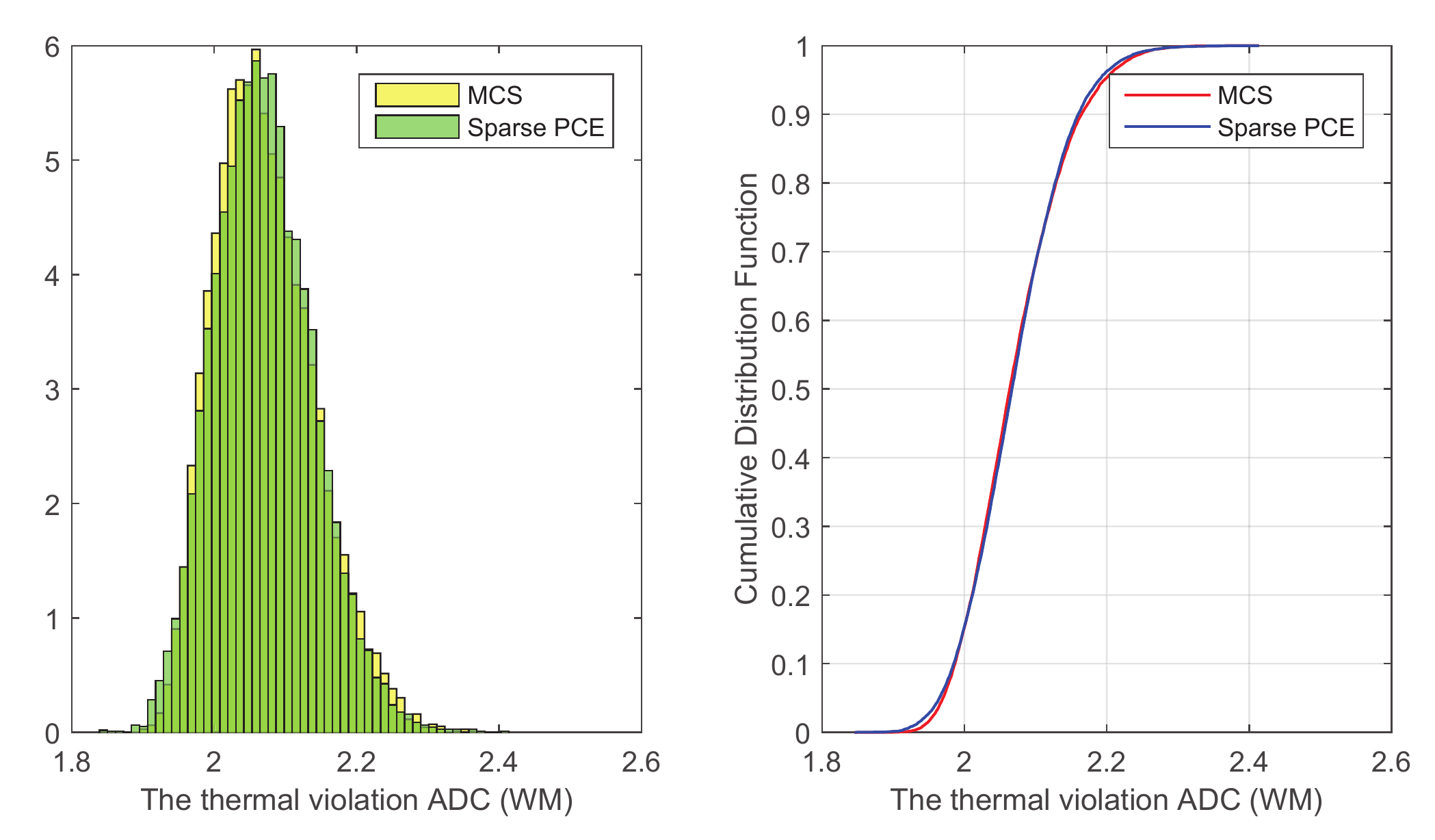}
\caption{The distribution of the thermal-violation ADC computed by MCS and sparse PCE. They are almost overlapped. To ensure 95\% confidence level, the thermal-violation ADC has to be decreased 4.51\% of the value calculated in deterministic system.}
\label{fig:adc_pdf_cdf_123}
\end{figure}

The thermal-violation ADC (the corresponding overall ADC in this case) without considering uncertainties is 2.063MW and the corresponding probability in CDF curve (Fig. \ref{fig:adc_pdf_cdf_123}) is 0.238, which means there is a risk of 23.8\% that operating the system to this margin will lead to unexpected thermal violation. To achieve 95\% confidence on operating the system securely, the thermal-violation ADC have to be decreased to 1.970MW, i.e., a 4.51\% reduction in ADC of the deterministic system. 

\begin{figure}[h]
\centering
\includegraphics[width=0.45\textwidth]{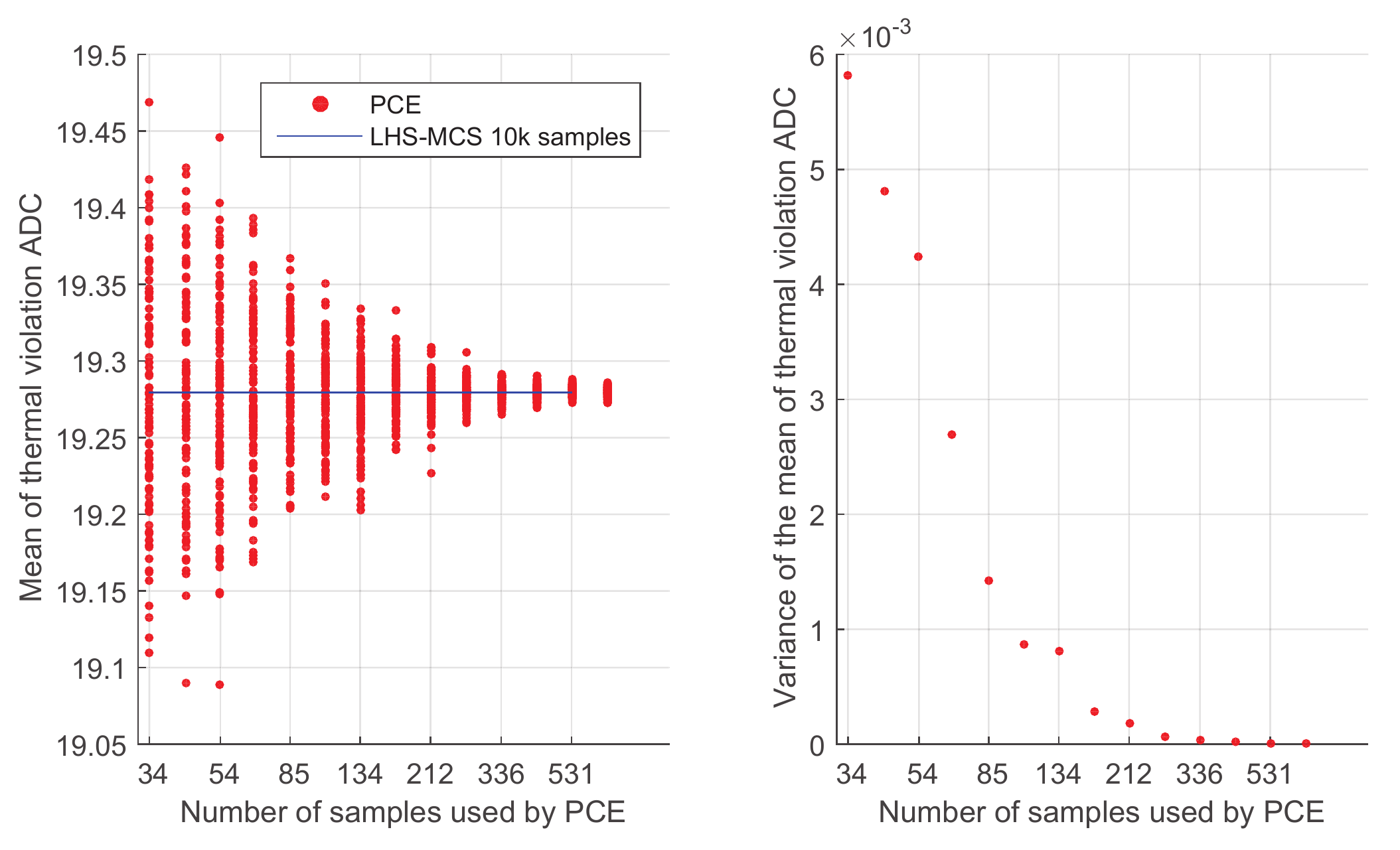}
\caption{The distribution of the mean of the thermal-violation ADC as the number of samples increases. Little improvement can be achieved by increasing sample size after it reaches 2.5 times of the number of random inputs (i.e., 335 in this case). }
\label{fig:adc_conv_rate_123}
\end{figure}

To verify the robustness of the proposed method, the convergence rate is traced by increasing the sample size from $0.25n$ to $5n$ ($n$ is the number of random inputs). For each sample size, 100 replications are generated and evaluated by the proposed method. As shown in Fig. \ref{fig:adc_conv_rate_123}, it is found that the statistics of ADC settle down to the values that computed by the LHS-based MCS using 10000 samples when the sample size reaches $2.5n$ (i.e., 335 in this case), after which little improvement can be achieved by increasing sample size. Similar results have been observed in the previous example, indicating a nice property of the sparse PCE that its simulation time grows linearly as the number of the random inputs increases.  

\section{Conclusion and Perspectives}
In this paper, we have proposed a formulation of probabilistic ADC for a general distribution network integrating various RES and load variations. 
A computationally efficient method to assess the probabilistic ADC is also developed, which combines the up-to-date sparse PCE and the continuation method.
The proposed method is able to handle a large number of correlated random inputs with diverse marginal probability distributions. Numerical studies show that the sparse PCE method can provide accurate estimations for the probabilistic ADC in terms of probability distribution and statistics with much less computational time ($\approx \frac{1}{30}$) than that of the MCS.
Moreover, the computational time grows linearly, rather than exponentially, as the number of random inputs increases.

The probabilistic ADC provides an intuitive yet significant picture regarding how the uncertainty brought about by the RES affects the delivery capabilities of a distribution network. It has been shown in our study that 4.51\% reduction in the ADC may be needed due to the randomness in the system to ensure the secure operation. This number may be even larger if the penetration of RES increases or their variances grow within a certain time period. This study points out that the increasing uncertainty to the system resulting from the growing penetration of RES needs to be carefully considered and its impact to the performances of a system has to be thoroughly investigated.

In the future, we plan to develop control measures to reduce the variance of ADC and thus increase the practical ADC by mitigating the violations at weak buses and branches. Appropriate control measures to reduce the variance of delivery capability will help compensate the impact of the randomness and thus enhance the overall security of modern power grids.



%




%

\ifCLASSOPTIONcaptionsoff
  \newpage
\fi

\begin{IEEEbiography}[{\includegraphics[width=1in,height=1.25in,clip,keepaspectratio]{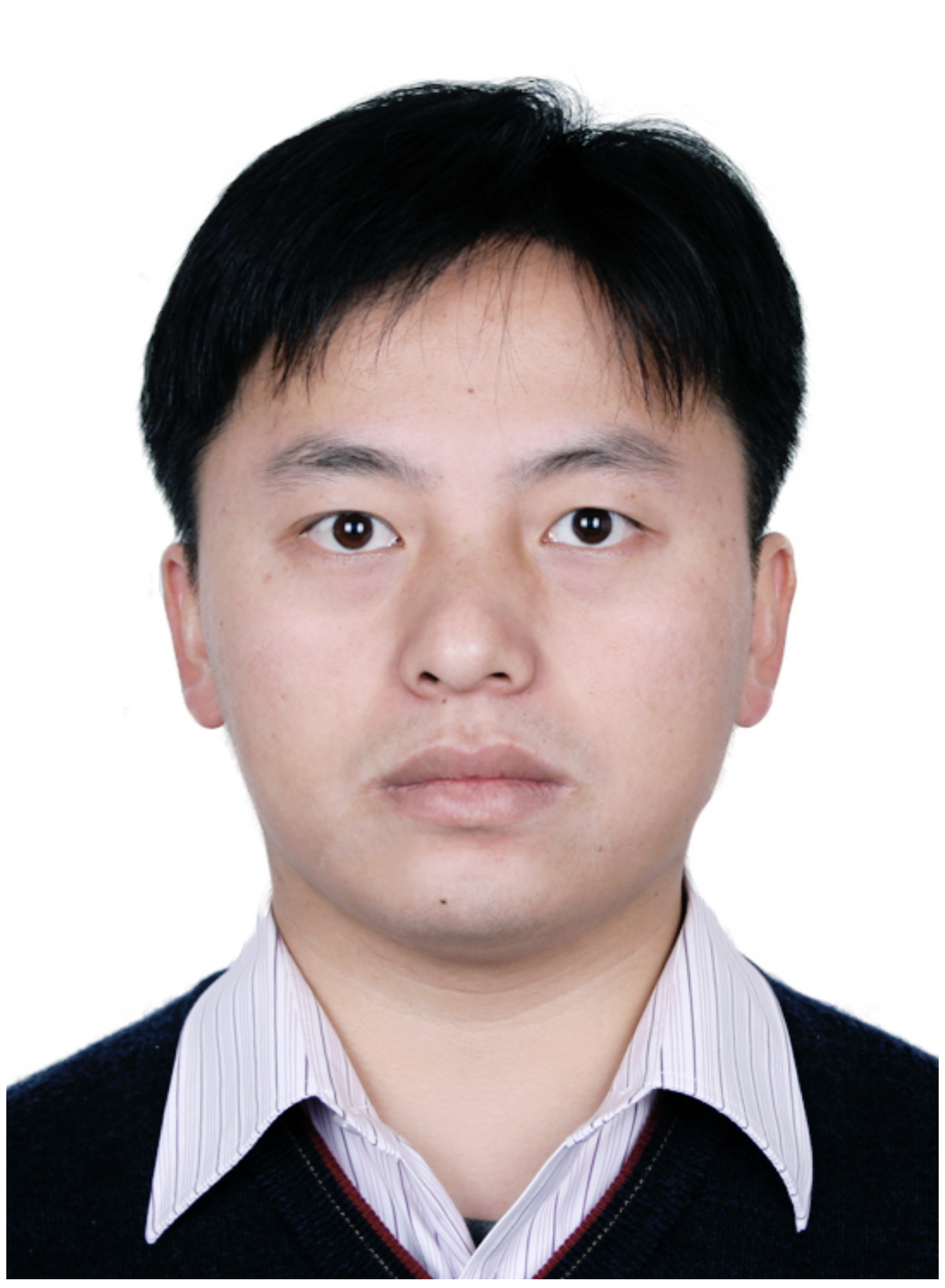}}]{Hao Sheng}
(M'14) is currently an postdoctoral fellow in the Department of Electrical and Computer Engineering at McGill University, Montreal, QC, Canada. 
He received the Ph.D. degree in the School of Electrical and Automation Engineering from Tianjin University, Tianjin, China, in 2014, the M.S. degree from Northeast Electric Power University, Jilin, China, in 2007 and the B.E. degree from North China Electric Power University, Baoding, China, in 2003, all in Electrical Engineering. From 2007 to 2012, he was affiliated with R\&D Centre of Beijing SiFang Automation Co., Ltd., Beijing, China, working on the development of PMU data-enhanced applications for Energy Management System (EMS) and Dynamic Security Assessment (DSA). From 2014 to 2017, he was a postdoctoral fellow in the School of Electrical and Computer Engineering at Cornell University, Ithaca, NY, USA. His research interests are in power system stability analysis and simulation, uncertainty quantification and control and their applications in power system static and dynamic security assessment.
\end{IEEEbiography}

\begin{IEEEbiography}[{\includegraphics[width=1in,height=1.25in,clip,keepaspectratio]{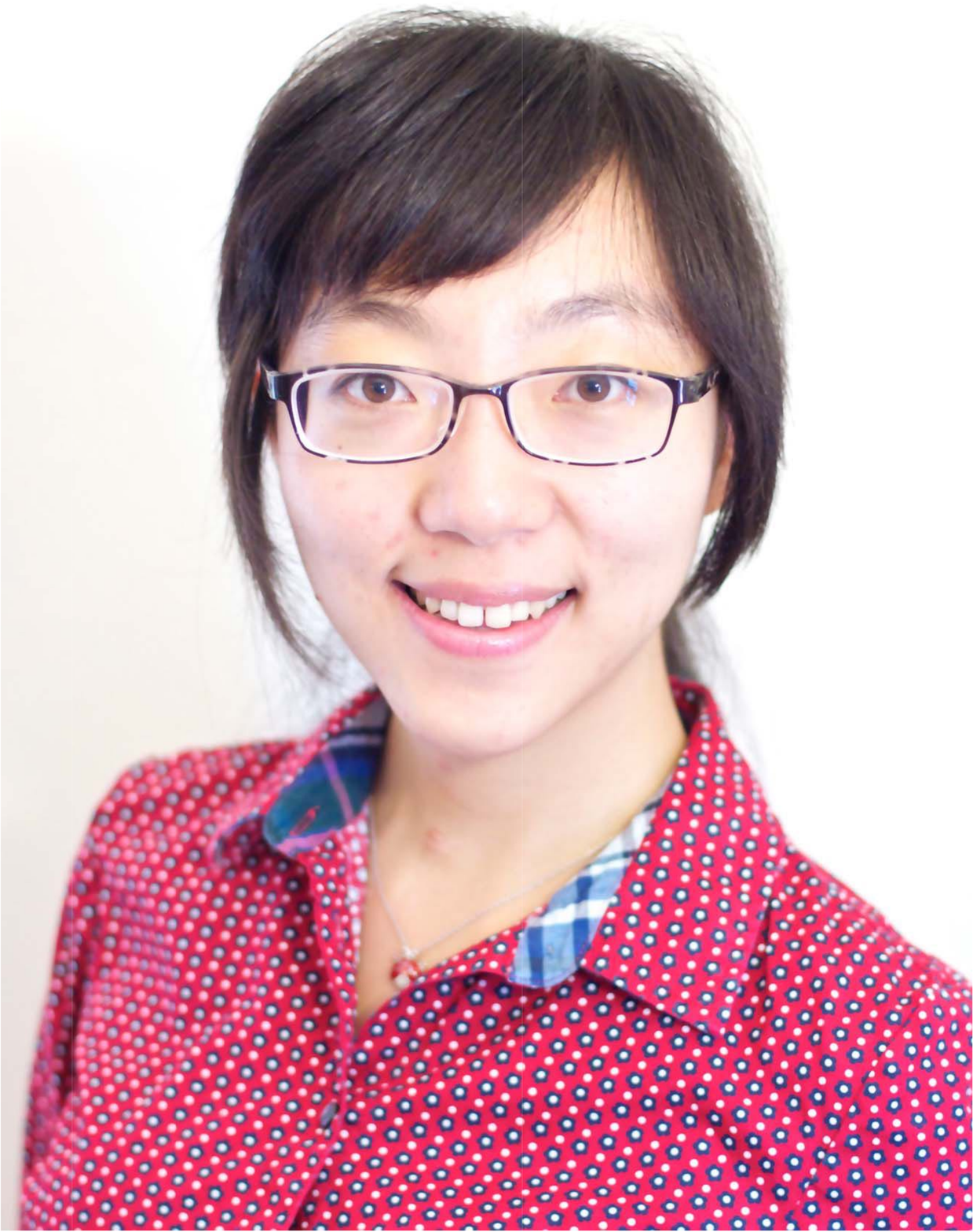}}]{Xiaozhe Wang}
is currently an Assistant Professor in the Department of Electrical and Computer Engineering at McGill University, Montreal, QC, Canada. She received the Ph.D. degree in the School of Electrical and Computer Engineering from Cornell University, Ithaca, NY, USA, in 2015, and the B.S. degree in Information Science \& Electronic Engineering from Zhejiang University, Zhejiang, China, in 2010. Her research interests are in the general areas of power system stability and control, uncertainty quantification in power system security and stability, and wide-area measurement system (WAMS)-based detection, estimation, and control.
\end{IEEEbiography}







\end{document}